\documentclass[prb,showpacs,twocolumn,preprintnumbers,amsmath,amssymb,floatfix]{revtex4}
\usepackage[dvips]{graphicx}
\usepackage[latin1]{inputenc}
\begin{document}
\draft

\newcommand{\Tb}  {[TbPc$_{2}$]$^{0}$}
\newcommand{\Dy}  {[DyPc$_{2}$]$^{0}$}
\newcommand{\etal} {{\it et al.} }
\newcommand{\ie} {{\it i.e.} }
\newcommand{\ip}{${\cal A}^2$ }

\hyphenation{a-long}

\title{Spin dynamics in rare earth single molecule magnets from $\mu$SR and NMR\\ in [TbPc$_{2}$]$^{0}$ and [DyPc$_{2}$]$^{0}$ }

\author{F. Branzoli$^1$, M. Filibian$^1$, P. Carretta$^1$, S. Klytaskaya$^2$ and M. Ruben$^2$}

\affiliation{$^1$ Department of Physics ``A.Volta", University of Pavia-CNISM, 27100 Pavia (Italy)}
\affiliation{$^2$ Institute of Nanotechnology, Karlsrhue Institute of Technology (KIT), 76344
Eggenstein-Leopoldshafen (Germany)}

\widetext

\begin{abstract}

The spin dynamics in [TbPc$_{2}$]$^{0}$ and [DyPc$_{2}$]$^{0}$ single molecule magnets have been investigated by
means of muon and nuclear spin-lattice relaxation rate measurements.  The correlation time for the spin
fluctuations was found to be close to 0.1 ms already at 50 K, about two orders of magnitude larger than the one
previously found in other lanthanide based single molecule magnets. In \Tb two different regimes for the spin
fluctuations have been evidenced: a high temperature activated one involving spin fluctuations across a barrier
$\Delta\simeq 880 K$ separating the ground and first excited states and a low temperature regime involving quantum
fluctuations within the twofold degenerate ground-state. In \Dy a high temperature activated spin dynamics is also
evidenced which, however, cannot be explained in terms of a single spin-phonon coupling constant.
\end{abstract}

\pacs {75.50.Xx, 76.75.+i, 76.60.-k} \maketitle

\narrowtext

Molecular nanomagnets have attracted major attention in the last decade owing to their possible technological
applicability as logic units in computers \cite{quantum} or in life sciences, as contrast agents \cite{CA}. So
far, the most studied molecular nanomagnets are the ones based on transition metal ions, which are characterized
by a twofold degenerate ground-state separated from the first excited states by a barrier induced by the magnetic
anisotropy \cite{Sessoli}. The stability of the ground-state and the applicability of these compounds above liquid
helium temperature (T) is nevertheless prevented by the small magnetic anisotropy which characterize transition
metal complexes \cite{Mn6}. In order to increase the anisotropy the synthesis of lanthanide based molecular
magnets has been envisaged and recently the double-decker phthalocyaninato lanthanide (Ln) complexes
[Pc$_{2}$Ln]$^{-}$ (Pc$\equiv$ C$_{32}$H$_{16}$N$_8$) have been recognized as potential single molecule magnets
(SMM) which could function above cryogenics temperatures \cite{Ishi1,Nano}. In fact, [Pc$_{2}$Tb]$^{-}$ and
[Pc$_{2}$Dy]$^{-}$ complexes are the first mononuclear lanthanide systems exhibiting a very slow relaxation and
quantum tunneling of magnetization \cite{Ishi1,Ishi2,Ishi5}. It has been recently observed \cite{Nos} by means of
nuclear magnetic resonance (NMR) that in [TbPc$_{2}$]$^{-}$TBA$^{+}\times$ N [TBA]Br, where N represents the
degree of dilution within the organic tetrabutylammonia (TBA) matrix, the characteristic correlation time
($\tau_c$) for the spin fluctuations becomes of the order of a $\mu$s at liquid nitrogen T,\cite{Nos} much longer
than in any other molecular magnet. These relatively long $\tau_c$ originates from the huge barrier separating the
twofold degenerate $|J= 6, m= \pm 6\rangle$ crystal field (CF) ground-state from the first excited $|J= 6, m= \pm
5\rangle$ state. In [TbPc$_{2}$]$^{-}$TBA$^{+}\times$ N [TBA]Br compounds, barriers as high as $\Delta\simeq 920$
K have been found \cite{Nos}. The CF splitting is determined both by [TbPc$_{2}$]$^{-}$ structure and by the
arrangement of the other molecules around it. In fact, upon varying N different values for $\Delta$ have been
estimated. Moreover, it has been observed that at temperatures $T\ll \Delta$ quantum fluctuations, related to the
tunneling between $m= \pm 6$ levels, arise.

In order to increase the stability of the ground-state the CF splitting has to be enhanced and the tunneling rate
lowered. In this work we have considered the effect of the one-electron oxidation of [Pc$_{2}$Ln]$^{-}$. The
resulting neutral [Pc$_{2}$Ln]$^{0}$ complex has an unpaired $\pi$ electron which resonates between the Pc
molecules, yielding a compression of the cage around Ln$^{3+}$ ion and possibly to an increase in the CF splitting
\cite{Ishi6,Ishi3,Ishi4}. Hereafter we present an investigation of the spin dynamics in two different neutral
compounds, \Tb and \Dy, by means of NMR and muon spin relaxation ($\mu$SR). The T dependence of $\tau_c$ has been
derived by means of  muon and nuclear spin-lattice relaxation rates and in \Tb\, found to be close to 0.1 ms,
already at 50 K, much longer than the one found in [TbPc$_{2}$]$^{-}$TBA$^{+}\times$ N [TBA]Br \cite{Nos}. The T
dependence of the relaxation rate is determined by the spin fluctuations induced by spin-phonon coupling.
Accordingly, information on the CF level splitting and on the spin-phonon coupling was derived.

\begin{figure}[h!]
\vspace{11.5cm} \includegraphics{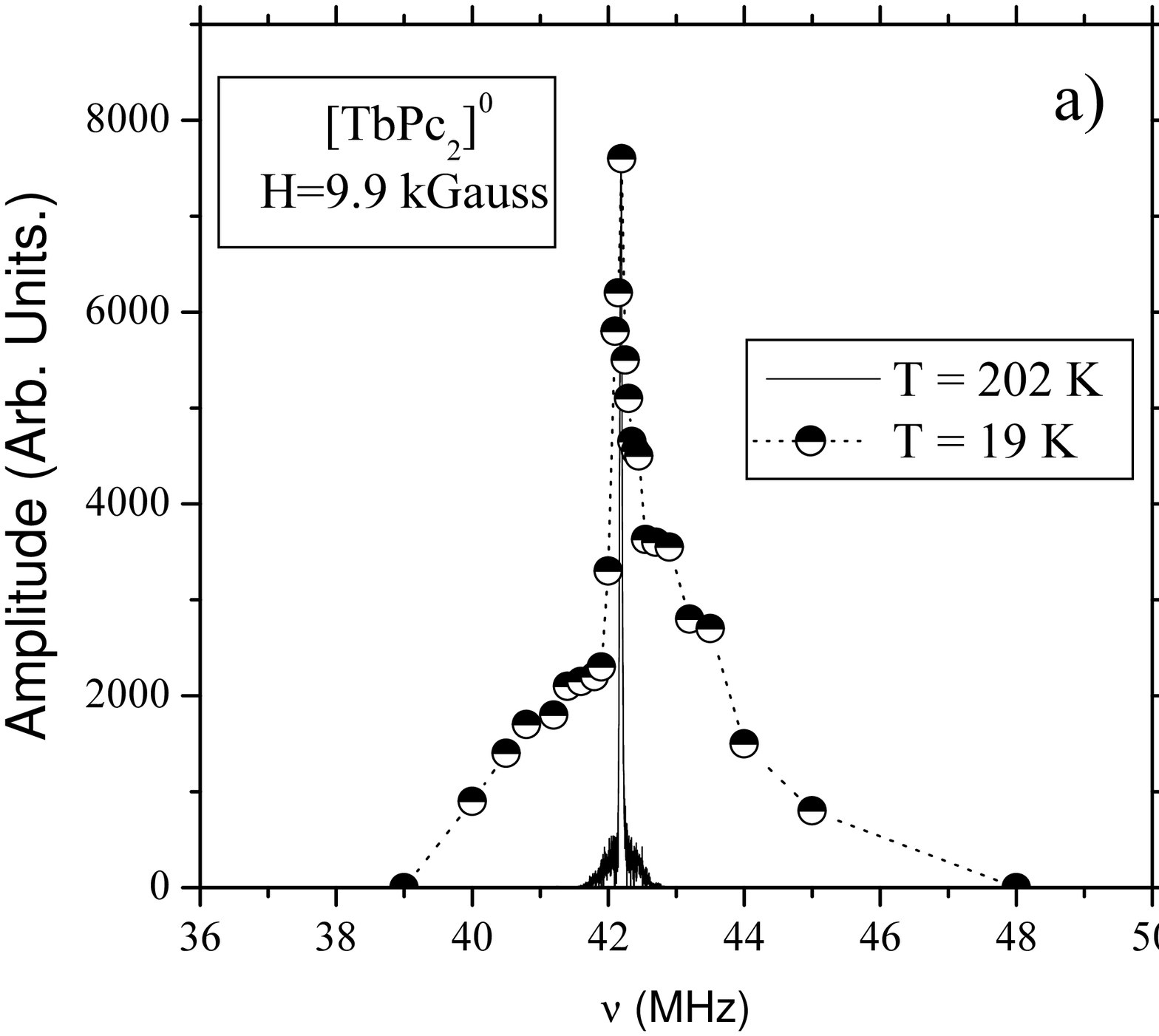} \includegraphics{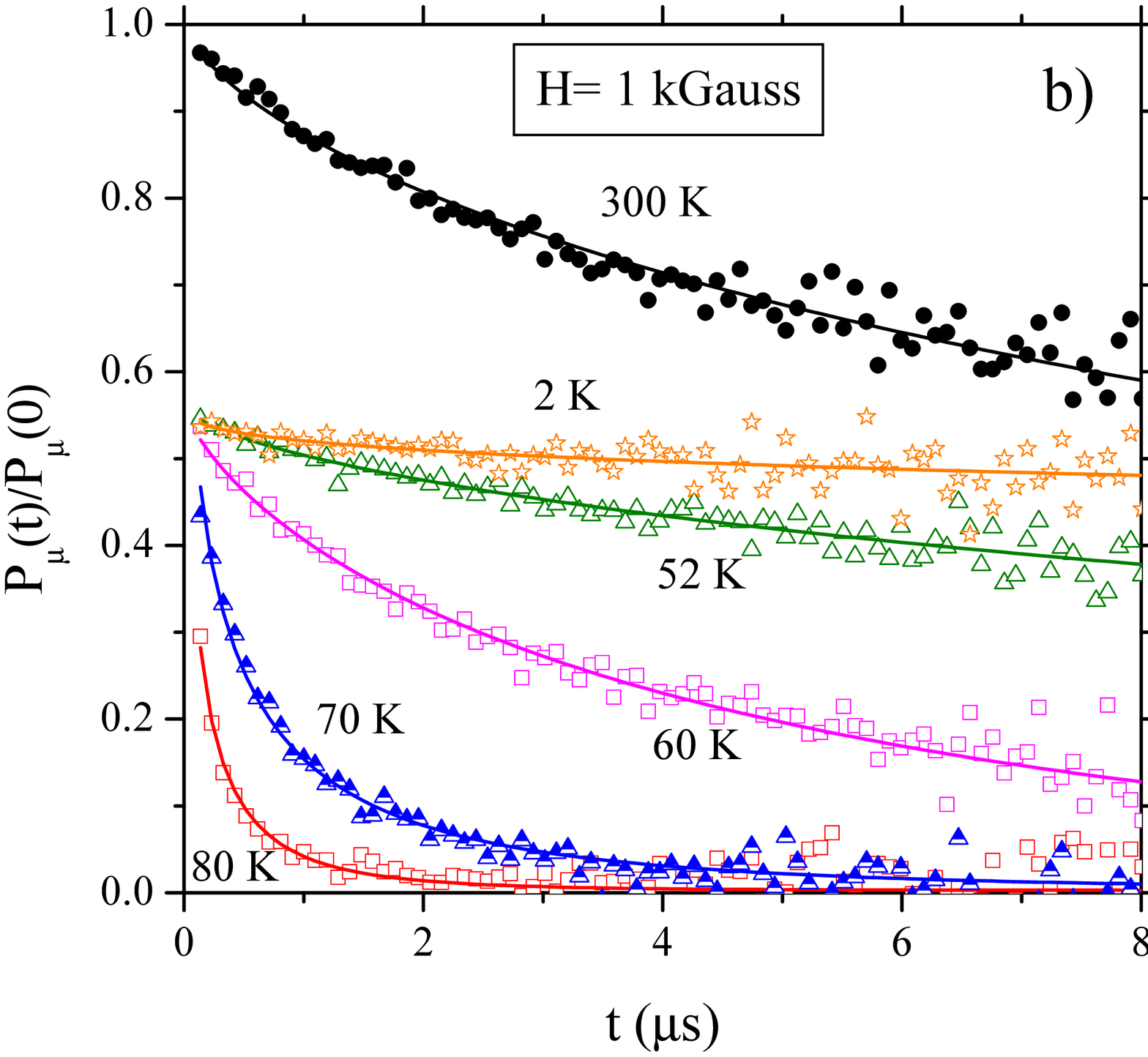} \caption{\textbf{a)} $^1$H NMR spectrum in [TbPc$_{2}$]$^{0}$ at $T=
19$ K (circles), compared to the much narrower spectrum obtained
  at $T = 202$ K (line). The linewidth at half intensity decreases from a few MHz at 19 K to about 37 kHz at 202 K.
  \textbf{b)} Time evolution of the muon polarization in [TbPc$_{2}$]$^{0}$ sample normalized to its value for $t\rightarrow 0$ at six
  selected temperatures.}
  \label{Fig1}
\end{figure}

\Tb and \Dy complexes  were synthesized according to the templating \cite{react} reaction of the phthalonitrile
precursor in boiling amyl alcohol in presence of DBU under inert atmosphere for 24 h. Treatment of [Ln(acac)$_3
\times$n H$_2$O] with phthalonitrile gave both forms, the anionic [Pc$_2$Ln]$^-$ and the neutral [Pc$_2$Ln]$^0$,
as observed in other earlier reports \cite{react}. The crude reaction mixture was presorbed on active
(H$_{2}$O-0\%) basic alumina oxide. This treatment helps to convert unstabilized anionic form to the neutral one.
Purification was carried out by column chromatography on basic alumina oxide (deactivated with 4.6\% H$_{2}$O,
level IV) with chloroform - methanol mixture (10:1) as eluent. Analytically pure powder samples were achieved by
additional radial chromatography on silica gel followed by recristallization from chloroform - hexane mixture.

$^1$H NMR measurements have been performed by using standard RF pulse sequences. While above 100 K the full NMR
line could be irradiated and the spectra obtained from the Fourier transform of half of the echo, below 40 K the
spectra became rather broad (Fig. 1a) and they could be derived from the envelope of  the echo amplitude upon
varying the irradiation frequency. This broadening is a clear evidence of the slowing down of the spin dynamics in
these SMM. Further information in this respect can be obtained from the study of the nuclear spin-lattice
relaxation rate $1/T_1$. $^1$H $1/T_1$ was estimated from the recovery of the nuclear magnetization after a
saturating RF pulse sequence. The recovery was found to be a stretched exponential, which evidences a distribution
of relaxation rates for the $^1$H nuclei, likely to be related to differences in the hyperfine coupling between
Ln$^{3+}$ and the 16 protons of each Pc molecule. The T dependence of $1/T_1$, reported in the inset of Fig. 2,
evidences a progressive increase upon cooling from RT down to $T \simeq 140$ K, while for $T\leq 40$ K $1/T_1$ is
found to progressively decrease. In the intermediate T range $140$ K$\geq T\geq$ 40 K the observation of $^1$H NMR
signal is prevented by the short relaxation times.

In order to explore the spin dynamics in these intermediate T range we have studied \Tb and \Dy by means of
$\mu$SR, which allows to measure much faster relaxation rates and to work at low fields, thus perturbing the SMM
only to a minor extent. Zero (ZF) and longitudinal field (LF) $\mu$SR experiments were carried out at ISIS pulsed
muon facility  on MUSR beam line. The background asymmetry contribution was found to be around 10\%  out of the
30\% total initial asymmetry, for both samples. The time decay of the muon polarization $P_{\mu}(t)$ in the
samples shows a different behaviour for temperatures above and below $T^*\simeq 90$ K, for \Tb, and $\simeq 60$ K,
for \Dy (Fig. 1b). Above $T^*$ the decay follows a stretched exponential behaviour $P_{\mu}(t)= A exp(-(\lambda
t)^{\beta})$, with an exponent $\beta\simeq 0.5$ and $A\simeq 20$\%.

\begin{figure}[h!]
\vspace{6.4cm} \includegraphics{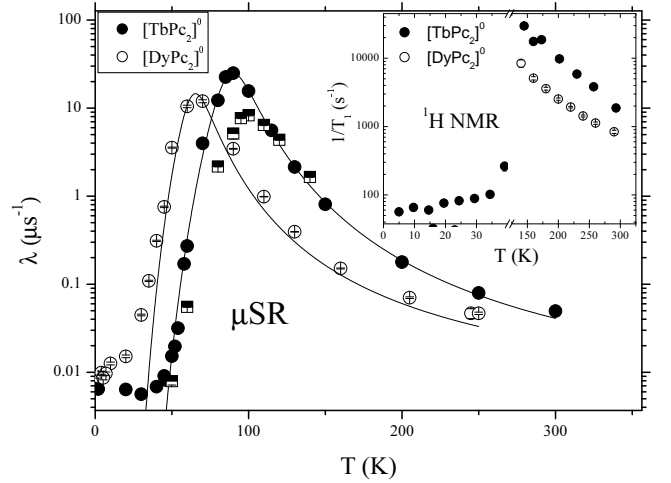}  \caption{T dependence of the
muon longitudinal relaxation rate in \Tb and \Dy for $H= 1000$ Gauss (circles) and for $H= 6000$ Gauss (squares).
The lines are the best fits according to Eq. 3. In the inset the $T$ dependence of $^1$H $1/T_1$ is reported for
the two compounds, for $H= 1$ Tesla.}
  \label{Fig2}
\end{figure}

Below $T^*$ a marked decrease of the initial asymmetry is found. This decrease has to be associated with the onset
of very low-frequency fluctuations. In particular, if the hyperfine field at the muon fluctuates with a
characteristic correlation time $\tau_c\gg 1/\gamma_{\mu}\sqrt{<\Delta h^2>}$, with $\gamma_{\mu}$ the muon
gyromagnetic ratio and $\sqrt{<\Delta h^2>}$ the root mean squared field distribution probed by the muons, the
decay of $P_{\mu}(t)$ is given by the static Kubo-Toyabe function \cite{Schenck}, which in zero-field is
characterized by a fast initial decay and by the subsequent recovery of the muon polarization to $A/3$. At ISIS
the initial fast decay cannot be detected, owing to the pulsed nature of the muon source and, accordingly, a loss
in the initial polarization is observed and only the long-time tail detected. At longitudinal fields $H\sim
\sqrt{<\Delta h^2>}$ an increase in the tail amplitude is expected. Since a LF of 1 kGauss (Fig. 1b) is found to
cause an increase of $P_{\mu}(t\rightarrow 0)/P_{\mu}(0)$ to $\alpha(H)\simeq 0.55$ for \Tb, one can estimate
$\sqrt{<\Delta h^2>}\simeq 650$ Gauss \cite{Schenck}. Thus, one has to expect that below $T^*$ Ln$^{3+}$ moments
are frozen over a timescale $(1/\gamma_{\mu}\sqrt{<\Delta h^2>})\sim 10$ ns. Still, even within this frozen
configuration low-energy excitations are present and drive muon spin-lattice relaxation. Hence, for $T< 80$ K, one
has that $P_{\mu}(t)= \alpha(H) A exp(-(\lambda t)^{\beta})$.

The T dependence of $\lambda$, derived by fitting the decay of the muon polarization according to the procedure
reported above is reported in Fig. 2. One notices a peak around 90 K for \Tb and around 60 K for \Dy, in agreement
with the NMR findings. The intensity of the peak is observed to scale with the inverse of the field intensity, a
situation found when the frequency of the fluctuations is close to Larmor frequency $\omega_L$.

\begin{figure}[h!]
\vspace{6cm} \includegraphics{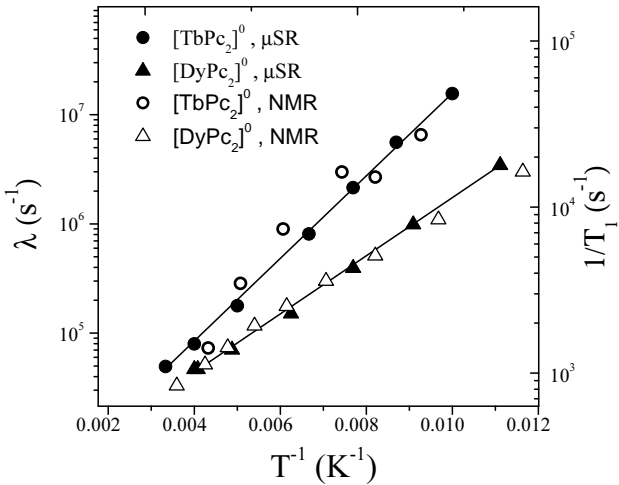}  \caption{Muon and $^1$H
spin-lattice relaxation rates are reported as a function of inverse T in a semi-logarithmic plot, in order to
evidence the activated behaviour for $T> T^*$. The lines are the best fits yielding $\Delta= 881$ K for \Tb and
$\Delta= 610$ K for \Dy. }
  \label{Fig3}
\end{figure}

From the T dependence of the spin-lattice relaxation rates it is possible to obtain information on the
characteristic correlation time for the spin fluctuations, on the energy barrier between the CF ground-state and
the first excited states and on spin-phonon coupling. In view of the energy difference between the muon (nuclear)
hyperfine levels and the $m$ levels of Ln$^{3+}$,  direct muon (nuclear) relaxation processes involving an
electron spin excitation are forbidden. The effective relaxation processes are indirect ones, involving a muon
(nuclear) spin flip without change in $|m|$. This is possible as the dipolar hyperfine Hamiltonian contains terms
coupling the transverse components of the hyperfine field $h_{x,y}$ to $J_z$. Owing to spin-phonon scattering
processes each CF level is characterized by a finite life-time ($\tau_m$) which yields a Lorentzian broadening and
$\lambda$ (or $1/T_1$) can be written as\cite{Borsa}
\begin{equation}
  {\lambda}\,\,\, \mathrm{or}\,\,\,\frac{1}{T_1}= \frac{\gamma^2\langle\Delta h_{\perp}^2\rangle}{Z} \sum_{m}
  \frac{e^{-E_m/T}\tau_m}{1+
\omega_L^2\tau_m^2}
\end{equation}
with $\langle\Delta h_{\perp}^2\rangle$ the mean-square amplitude of the hyperfine field fluctuations, $E_m$ the
eigenvalues of the CF levels and $Z$ the corresponding partition function. The life-time for the $m$ levels can
also be expressed in terms of the transition probabilities $p_{m, m\pm 1}$ between $m$ and $m\pm 1$ in the form
$1/\tau_m= p_{m, m+1} + p_{m, m-1}$. Since the life-times are mainly determined by spin-phonon scattering
processes those can be expressed in terms of the CF eigenvalues \cite{Villain}
\begin{equation}
p_{m, m\pm 1}= C \frac{(E_{m\pm 1}-E_m)^3}{e^{\beta (E_{m\pm 1}-E_m)} -1}
\end{equation}
with $C$ spin spin-phonon coupling constant and $\beta= 1/T$. Since in \Tb \, $\Delta= E_{m=\pm 5}- E_{m=\pm 6}\gg
k_BT$ over all the explored T range, Eq. 2 can be simplified in the form
\begin{equation}
     {\lambda}\,\,\, \mathrm{or}\,\,\,\frac{1}{T_1}=
     {{\gamma^{2}\langle\Delta h_{\perp}^2\rangle}\over {2}}{2\tau_c\over 1+\omega^{2}_L \tau^{2}_c}
  \,\,\, ,
\end{equation}
with $\tau_c= exp(\Delta/T)/(C \Delta^3)$. From the amplitude of $\lambda$ at $T^*$ one can estimate a  root mean
square amplitude for the fluctuating field at the muon in \Tb $\sqrt{\langle\Delta h^{2}_\perp\rangle}\simeq 770$
Gauss. This value is close to the one of the field distribution giving rise to the fast initial relaxation of the
muon polarization, described by Kubo-Toyabe function. By fitting the data with Eq. 3 it is possible to estimate a
spin-phonon coupling constant $C\simeq 3000$ Hz/K$^3$, of the same order of magnitude of that found in other SMM
\cite{Nos,Borsa}. It is noticed that for $T> 100$ K, $\tau_c\omega_L\ll 1$ and from Eq. 3 one has $\lambda\propto
exp(\Delta/T)$. In fact, by plotting either $1/T_1$ or $\lambda$ data vs. $1/T$ in a semi-logarithmic scale (Fig.
3), one finds a nice linear behaviour, consistent with a value of $\Delta\simeq 880$ K.

On the basis of Eq. 3, in the light of the data reported in Fig. 2, it is possible to derive the T dependence of
$\tau_c$ for \Tb.  In \Tb\, $\tau_c$ shows two regimes (Fig. 4): a high T activated one and a low T one, for $T<
50$ K, where the correlation time is constant. The high T behaviour describes spin fluctuations among $m=\pm 6$
and $m=\pm 5$ levels, while the low T trend rather signals tunneling processes among $m= +6$ and $m= -6$ levels.
The total fluctuation rate is given by the sum of the fluctuation rates associated with each process and
accordingly one has
\begin{equation}
    {1\over \tau_c}=({1\over \tau_c})_{act}+({1\over \tau_c})_{t}= C\Delta^3 e^{-\Delta/T} + ({1\over \tau_c})_{t}
\end{equation}
with a tunneling rate $(1/\tau_c)_{t}\simeq 90 \mu s$. It is noticed (Fig. 4) that the $T$-dependence of $\tau_c$
can be reproduced very well with the above equation. It is now useful to compare NMR and $\mu$SR results in \Tb
with AC susceptibility ones \cite{Ishi1,Ishi3,Ishi4,Ishi6}. In fact, also these measurements evidence a high T
activated behaviour with energy barriers which appear to be only slightly smaller than the ones found here. On the
other hand, the role of tunneling processes in the AC susceptibility measurements has not been addressed for the
Pc$_2$Ln complexes  and just the frequency dependence of the AC susceptibility peak discussed \cite{Ishi3,Ishi6}.
Nevertheless, we remark that also the imaginary part of the AC susceptibility is characterized by a low T
flattening which might originate from tunneling processes \cite{Ishi3,Ishi6}. Still, in comparing these techniques
one should consider that local techniques as NMR and $\mu$SR are sensitive to the spin fluctuations of each
molecule even if they do not yield a net variation in the macroscopic magnetization, namely are silent in the AC
susceptibility \cite{Ni10}.

\begin{figure}[h!]
\vspace{6.2cm} \includegraphics{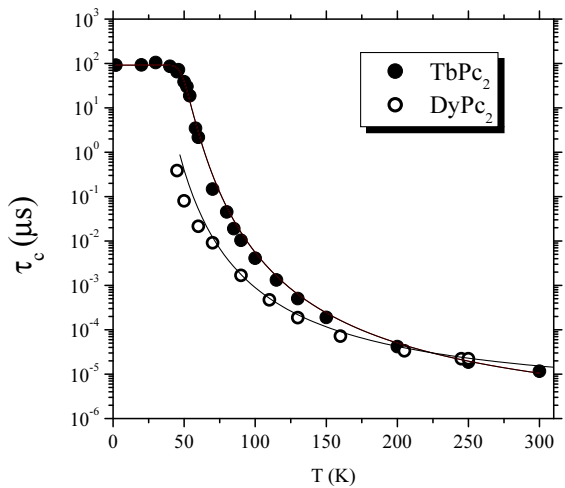}  \caption{T dependence of
the correlation time for the spin fluctuations in \Tb and \Dy (for $T\geq T^*$) , derived from $\lambda$ data
reported in Fig. 2, on the basis of Eq. 3. The solid lines are the best fits according to Eq. 4. }
  \label{Fig4}
\end{figure}

Now, if one considers the T dependence of $\lambda$ in \Dy (Fig. 2), one notices a behaviour quite similar to the
one found in \Tb, with a maximum around 60 K characterized by a slightly lower amplitude. This decrease in the
amplitude has to be associated with a lower hyperfine coupling, corresponding to  a  root mean square amplitude
for the fluctuating field at the muon $\sqrt{\langle\Delta h^{2}_\perp\rangle}\simeq 590$ Gauss. Since Dy$^{3+}$
magnetic moment is slightly larger than Tb$^{3+}$ one, this decrease rather indicates a slightly larger average
distance between the rare earth ion and the muon in \Dy. In view of the similarity with \Tb one could try to fit
also \Dy $\lambda(T)$ data with Eq. 3. One finds a good fit only for $T> 40$ K and one can estimate a barrier for
the spin fluctuations around $610$ K (Fig. 3) and a spin-phonon coupling constant $C\simeq 1780$ Hz/K$^3$.
However, one notices that for $T< 40$ K the data are not so well reproduced. In fact, for \Dy it is likely that
the two lowest doublets are separated by an energy barrier in the tens of K range \cite{Ishi1} and that the high
barrier derived from $\lambda(T)$ corresponds to the one between the first and second excited doublets $|J, \pm
m\rangle$. Hence, in order to appropriately fit the data one should resort to Eq. 1. However, even by allowing
different possible values for the CF splittings a poor fit of \Dy data is obtained. It is likely that one should
consider that the spin-phonon coupling constant is not the same for the scattering processes yielding high and low
energy transitions.

In conclusion, by means of $\mu$SR and NMR spin-lattice relaxation rates we have derived the T dependence of the
characteristic correlation time for the spin fluctuations in \Tb and \Dy SMM. In \Tb, although the barrier between
the ground and first excited states is close to the one of [TbPc$_2$]$^-$ SMM diluted in a TBA matrix \cite{Nos},
$\tau_c$ is found to be almost two orders of magnitude larger for $T< 50$ K. This suggests that in the neutral SMM
the structural arrangement around the SMM itself \cite{Ruben} leads to an enhancement of the CF symmetry and to a
decrease of the quantum fluctuations within the twofold degenerate ground-state. Thus, it appears of major
relevance, in order to enhance $\tau_c$ and to be able to employ SMM at economically affordable temperatures, to
carefully investigate the effect of the modifications in the SMM coordination on the spin dynamics.

Useful Discussions with F. Borsa are gratefully acknowledged. The research activity in Pavia was supported by
Fondazione Cariplo 2008-2229 research funds.



\end{document}